\documentclass[doublecol]{epl2}
\usepackage{amsmath,amsfonts,amssymb,bm}
\usepackage{graphicx,eucal,psfrag}

\title{Dynamical non-ergodic scaling in continuous finite-order
quantum phase transitions}

\author{S. Deng \inst{1}
\and G. Ortiz \inst{2} \and L. Viola \inst{1} } \shortauthor{S.
Deng \etal}

\institute{
  \inst{1} {Department of Physics and Astronomy, Dartmouth
College, 6127 Wilder Laboratory, Hanover, NH 03755, USA} \\
  \inst{2} Department of Physics, Indiana University, Bloomington,
IN 47405, USA }

\date{\today}


\abstract{We investigate the emergence of universal dynamical scaling
in quantum critical spin systems adiabatically driven out of
equilibrium, with emphasis on quench dynamics which involves
non-isolated critical points ({\em i.e.}, critical regions) and cannot
be {\em a priori} described through standard scaling arguments nor
time-dependent perturbative approaches.  Comparing to the case of an
isolated quantum critical point, we find that non-equilibrium scaling
behavior of a large class of physical observables may still be
explained in terms of equilibrium critical exponents. However, the
latter are in general non-trivially path-dependent, and detailed
knowledge about the time-dependent excitation process becomes
essential. In particular, we show how multiple level crossings within
a gapless phase may completely suppress excitation depending on the
control path. Our results typify non-ergodic scaling in continuous
finite-order quantum phase transitions.}

\begin{document}
\maketitle

The response of a physical system to external probes is an invaluable
technique for unveiling the system's properties. If the probe is
dynamic, so that the Hamiltonian becomes explicitly time-dependent,
the system is forced out of equilibrium -- a subject of prime
practical importance which can soon prove full of challenges and
surprises. In particular, understanding and manipulating the dynamics
of zero-temperature quantum phase transitions (QPTs) \cite{Sachdev} in
matter has a broad significance across fields as diverse as quantum
statistical mechanics, material science, quantum information
processing, and cosmology. The extent to which universal quantum
scaling laws persist out of equilibrium and encode information about
the equilibrium phase diagram is the topic of this work.

As early as 1970, Barouch and coworkers \cite{Barouch} studied the
time-dependent $T=0$ magnetization of the anisotropic XY chain,
and showed that equilibrium is not reached at the final evolution
time. This {\it non-ergodic} behavior was later confirmed for
other physical observables \cite{Joseph}, and the analysis
extended to the case where the system is driven across its quantum
critical point (QCP) by changing a control parameter $\lambda(t)$
({\em e.g.}, the magnetic field along the $z$-axis) in time with
constant quench rate $\tau
>0$.  The emergence of non-equilibrium scaling, however, was not
discussed. An important step was taken in Ref. \cite{Zurek1}, starting
from the observation that irrespective of how slowly the quench
occurs, adiabaticity is lost in the thermodynamic limit at a
``freeze-out'' time $(t_c-\hat{t})$ before the QCP is crossed.  This
yields a power-law prediction for the final density of excitations,
$n_{\text{ex}}(t_{\text{fin}}) \sim \xi^{-1}(\hat{t}) \sim
\tau^{-\ell}$, where the {\em non-equilibrium critical exponent}
$\ell=d \nu/(\nu z+1)$ is solely determined by the equilibrium
correlation length ($\xi$) the dynamic critical exponents of the QCP
($\nu$ and $z$, respectively), and the spatial dimension, $d$.  While
it is suggestive to realize that defect formation is a manifestation
of broken ergodicity in Barouch's sense \cite{Joseph}, continuous
experimental advances in systems ranging from ultracold atomic gases
to quantum magnets \cite{exp} demand the applicability of the above
{\em Kibble-Zurek scaling} (KZS) to be carefully scrutinized, and the
potential for more general {\it non-ergodic} scaling to be
explored. How much information on the equilibrium physics is needed
for reliable scaling predictions to be possible?

The KZS for linear quenches across an {\em isolated} QCP
separating two gapped phases has been confirmed by now for a
variety of control schemes in one-dimensional (1D) models
\cite{Jacek,Anatoli,Cherng}, including QCPs of topological nature
\cite{Mondal} and noisy driving fields \cite{Fubini} --
generalizations to repeated \cite{Victor} and non-linear quenches
\cite{Sen,Sen2} having also been established. Leaving aside the
case of {\em disordered} quantum systems, where marked deviations
from power-law behavior may be witnessed \cite{random}, the
possibility of genuinely {\em non-adiabatic} scaling in
low-dimensional clean systems, whereby non-zero excitation
persists (unlike KZS) for $\tau \rightarrow \infty$ in the
thermodynamic limit, has been pointed out in \cite{Anatoli}.
Likewise, critical dynamics in the presence of {\em non-isolated}
QCPs reveals a rich landscape. The need to modify the KZS by
replacing $d$ with the ``co-dimension'' $m$ of the relevant
critical (gapless) surface has emerged through a study of the 2D
Kitaev model \cite{Sengupta}.  Evidence of non-KZS has also been
reported for quenches which originate within an extended quantum
critical region \cite{Pellegrini}, cross a multi-critical point
\cite{Victor2,Sen2}, or steer the system along  a gapless critical
line \cite{Div}.

A main purpose of this work is to develop a theory and
understanding of non-ergodic scaling for generic (power-law)
quenches along critical regions. To achieve this goal two new
notions are introduced, which are both {\em path-dependent}: one
is the concept of a {\it dominant critical point} to establish
scaling along a critical path, and the other a {\it mechanism of
cancellation} of excitations.  Besides elucidating several results
recently reported in the literature, our analysis indicates that
details on how different modes of excitation are accessed
throughout the quench process are crucial. We consider several
different scenarios within a unifying illustrative testbed, the 1D
anisotropic XY model in a transverse alternating magnetic field
\cite{deng}. In particular, we push beyond the KZS domain --
notably, by investigating quenches that involve a continuous
Lifshitz QPT to a gapless phase.  We also revisit the standard KZS
and clarify how, for arbitrary continuous QPTs, it can be
accounted for by the {\em iterative adiabatic renormalization}
approach of Berry \cite{Berry}, as long as two gapped quantum
phases are involved. Most importantly, we find that universal
dynamical scaling is obeyed by a large class of extensive physical
observables {\em throughout the quench dynamics}, a result with
practical implications in the experimental detection of
non-ergodic scaling.

%%%%%%%%%%%%%%%%%%%%%%%%%%%%%%%%%%%%%%%%%%%%%%%%%%%%%%%%%%%%%%%%%%%%%%
\vspace*{0.5mm} {\bf Model Hamiltonian.--} The spin-$1/2$
anisotropic XY model in a transverse alternating field is defined
by \cite{deng}
\begin{eqnarray*}
H = - \hspace*{-0.8mm}\sum_{i=1}^N \hspace*{-0.8mm} \Big\{
\frac{1+\gamma}{2} \sigma_x^i \sigma_x^{i+1} \hspace*{-0.8mm} +
\frac{ 1-\gamma }{2} \sigma_y^i \sigma_y^{i+1} - \hspace*{-0.7mm}
[h -(-)^i \delta ] \sigma_z^i \Big\},
\end{eqnarray*}
\vspace*{-5mm}
\begin{equation}
\label{Ham}
\end{equation}
where periodic boundary conditions are assumed, that is,
$\sigma_\alpha^i \equiv \sigma_\alpha^{i+N}$. Here, $\gamma \in
[0,1]$, $h, \delta \in [-\infty, \infty]$, are the anisotropy in
the XY plane, and the uniform and alternating magnetic field
strength, respectively.  This model can be exactly solved by
following the steps outlined in \cite{Okamoto,deng}.  The
Hamiltonian (\ref{Ham}) rewrites as $H=\sum_{k \in K_+}
\hat{H}_k=\sum_{k \in K_+} {A}_k^{\dag} {H}_k {A}_k$, where $K_+ =
\{\pi/N,3\pi/N,\ldots, \pi/2-\pi/N \}$ specifies allowed momentum
values, and ${A}_k^\dag=(a_k^\dag, a_{-k}, b_k^\dag, b_{-k})$ is a
vector operator, with $a_k^\dag$ ($b_k^\dag$) denoting canonical
fermionic operators that create a spinless fermion with momentum
$k$ for even (odd) sites. Diagonalization of the reduced $4 \times
4$ Hamiltonian matrix $H_k$ further yields a collection of
non-interacting quasi-particles,
$$H=\sum_{k\in K_+}^{n=1, \ldots,4} \epsilon_{k,n}
N_{k,n},$$ \noindent in terms of an appropriate number operator
$N_{k,n}$ for mode $(k,n)$.  Assuming that $n$ labels bands in
increasing energy order, only $\epsilon_{k,1},\epsilon_{k,2} \leq
0$ bands are occupied at $T=0$, with an excitation gap
$\Delta_k=\epsilon_{k,3}-\epsilon_{k,2}$ being given
by:\begin{eqnarray} \Delta_k (\gamma, h, \delta) &=&
4\Big[\, h^2+\delta^2 + \cos^2 k+\gamma^2 \sin^2 k \label{gap} \\
&- &2\sqrt{h^2\cos^2k+\delta^2(h^2+\gamma^2\sin^2k)}
\,\Big]^{1/2}. \nonumber \end{eqnarray}

\begin{figure}
\psfrag{x}{$\hspace*{-1mm} {h}$}
\psfrag{y}{$\hspace*{-1mm}{\delta}$}
\includegraphics[width=\linewidth]{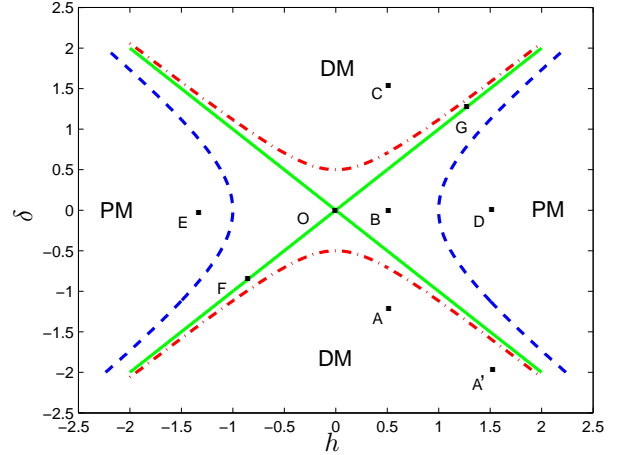}
\vspace*{-5mm} \caption{\label{pd00}(Color on-line) Phase diagram
of the alternating spin chain, Eq. (\ref{Ham}). Dashed (blue) and
dashed-dotted (red) lines define the phase boundaries for
$\gamma=0.5$, the enclosed area corresponding to the FM phase.
Dashed (blue) and solid (green) lines correspond to $\gamma=0$,
the enclosed area being the SF gapless phase. }
\end{figure}

Quantum phase boundaries are determined by the equations $h^2 =
\delta^2+1$; $\delta^2 = h^2+\gamma^2$.  The phase diagram with both
$\gamma=0.5$ and $\gamma=0$ is shown in Fig.~\ref{pd00}.  Quantum
phases corresponding to disordered paramagnetic (PM) and dimer (DM)
behavior emerge as depicted for arbitrary $\gamma$. For $\gamma >0$,
ferromagnetic order (FM phase) develops in the center of the phase
diagram, whereas for the isotropic chain a superfluid (SF) phase with
a gapless spectrum and nonbroken $U(1)$-symmetry emerges.  Finite-size
analysis reveals that this model supports four distinct universality
classes: (i) When $\gamma > 0$, generic QCPs belong to the $d=2$ Ising
universality class with critical exponents $\nu=1,z=1$.  Different
critical behavior occurs at $(h \rightarrow 0, \delta = \pm\gamma)$
and $(h = \pm1, \delta \rightarrow 0)$, where weak singularities in
the ground-state energy develop (4th-order QCPs \cite{deng}), and
$\nu=2, z=1$, corresponding to the alternating universality class;
(ii) When $\gamma=0$, generic QCPs on the boundary lines belong to the
Lifshitz universality class, with critical exponents $\nu=1/2, z=2$.
Different critical behavior still occurs at ($h=\pm 1, \delta
\rightarrow 0$), where now $\nu=1,z=2$.  Furthermore, Ising critical
exponents are recovered while approaching the point $(h=0, \delta=0)
\equiv O$ along every path other than $(\delta=0, h \rightarrow 0)$
(when there is no QCP). In what follows, we focus on quenching schemes
where $h$ and $\delta$ are individually or simultaneously varied with
time. We address separately different representative scenarios.

\vspace*{0.5mm} {\bf Quenching across an isolated critical
point.--} Suppose first that the system is {\em linearly} quenched
across an isolated (non-multicritical) QCP that separates two
gapped phases upon changing a single control parameter as $\delta
\lambda(t)=\lambda(t)-\lambda_c = (t-t_c)/\tau$, where $t \in
[t_{\text{in}}, t_{\text{fin}}]$. Without loss of generality we
may assume that the system becomes critical at $t_c=0$. For finite
$N$, the exact time-evolved many-body state $|\psi(t)\rangle$ may
be determined from numerical integration of the Schr\"odinger
equation with Hamiltonian $H(t)$, subject to
%the initial condition
$|\psi(t_{\text{in}})\rangle = |\psi_{GS}(t_{\text{in}})\rangle$, the
latter being the ground state of $H(t_{\text{in}})$.  The final
excitation density $n_{\text{ex}}(t_{\text{fin}})$ may then be
computed from the expectation value of the instantaneous
quasi-particle number operator over $|\psi(t)\rangle$.
%$n_{\text{ex}}(t_{\text{fin}}) = 1/N \langle \psi(t_{\text{fin}})|
%\sum_{k;n} N_{k,n}(t) |\psi (t_{\text{fin}})\rangle$.
Provided that the quench rate $\tau$ belongs to the appropriate
range\footnote[1]{That $\tau \geq \tau_{\text{min}}$ follows from
standard adiabaticity requirements away from criticality,
$\tau_{\text{min}}$ $ \sim 1/ [\text{min}_{t \in
[t_{\text{in}},t_{\text{fin}}]} \text{Gap}(H(t))]^{2}$. The
existence of a finite upper bound $\tau_{\text{max}}$ follows from
the fact that if $\tau$ is arbitrarily large, a {\em finite}
system never enters the impulsive regime, if the size-dependent
contribution to the gap dominates over the control-dependent one.
From scaling analysis under the assumption that the gap closes
polynomially as $N^{-z}$, we estimate $\tau_{\text{max}} \sim
N^{(\nu z+1)/\nu}$.}, KZS is found to hold irrespective of the
details of the QCP and the initial (final) quantum phase, in
particular for both 2nd and higher-order QPTs, and independent of
the path direction:
$$ n_{\text{ex}}^{\text{Ising}} (t_{\text{fin}})\sim \tau^{-1/2}\,,
\;\;\;\;\; n_{\text{ex}}^{\text {Alternating}} (t_{\text{fin}})
\sim \tau^{-2/3}\,.
$$

\begin{figure}
\includegraphics[width=\linewidth]{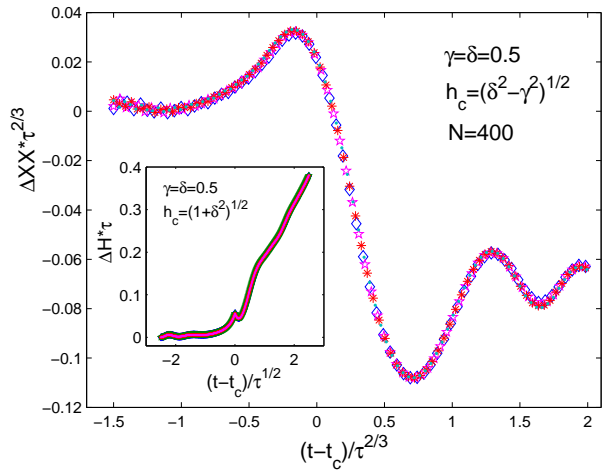}
\vspace*{-5mm} \caption{\label{tpu} (Color on-line) Dynamical
scaling under a magnetic field quench. Main panel: excess
nearest-neighbor spin correlation per particle, $\Delta XX$, vs
rescaled time for the alternating universality class from
numerical integration of the Schr\"{o}dinger equation. Inset:
excess energy per particle, $\Delta H$, vs rescaled time for the
Ising universality class from first-order adiabatic
renormalization. }
\end{figure}

While the excitation density is an accurate measure of the loss of
adiabaticity in exactly-solvable models, identifying manifestations of
the KZS in quantities that can be more directly accessible in
experiments and/or meaningful in more general systems is essential.
Remarkably, numerical results indicate that scaling behavior holds
{throughout the quench process} for a large class of physical
observables, provided that the excess expectation value relative to
the instantaneous ground state is considered \cite{deng}. That is,
\begin{eqnarray}
\Delta {\cal O}(t) &\equiv & \langle \psi(t)| \, {\cal O}\,
|\psi(t)\rangle - \langle \psi_{GS}(t) |\, {\cal O}\, |\psi_{GS}
(t) \rangle \nonumber \\ &=& \tau^{-(\nu +\beta)/(\nu z+1)}F_{\cal
O}\Big(\frac{t-t_c}{\hat{t}}\Big)\,, \label{scaling}
\end{eqnarray}
where $\beta$ is a scaling exponent determined by the physical
dimension of ${\cal O}$ and $F$ is an observable-dependent scaling
function. For instance, under a quench of the magnetic field strength,
$h$, the magnetization per site, $M_z=(\sum_{i=1}^N \sigma_z^i)/N$,
obeys dynamical scaling of the form $\Delta M_z(t) =\tau^{(-\nu-\nu
z+1)/(\nu z+1)}G({(t-t_c)}/{\hat{t}})$, whereas the nearest-neighbor
spin correlator along the $x$-direction, $XX=(\sum_{i=1}^N \sigma_x^i
\sigma_x^{i+1})/N$, obeys dynamical scaling of the form $\Delta XX(t)
=\tau^{-\nu/(\nu z+1)} W({(t-t_c)}/{\hat{t}})$, for appropriate
scaling functions $G$ and $W$, respectively -- see Fig.~\ref{tpu}.

The fact that the system becomes gapless at a single instant $t_c$
suggests to seek an explanation of the above results based on the fact
that $\dot{\lambda}(t)=1/\tau$ is a small parameter.  While a similar
strategy has been implemented in \cite{Anatoli}, our emphasis is on
providing a firm theoretical foundation and further highlighting
important assumptions. By a suitable parametrization, the relevant
time-dependent Hamiltonian may be written as
$H(t)=H_c+[\lambda(t)-\lambda_c] H_1= H_c + (t-t_c)/\tau H_1$, with
$H_c$ quantum-critical in the thermodynamic limit. Let $\{E_m(t)\}$
and $\{|\psi_m(t)\rangle \}$ denote the snapshot eigenvalues and
(orthonormal) eigenvectors of $H(t)$, where $|\psi_0(t)\rangle \equiv
|\psi_{GS} (t)\rangle$ and assume that: (i) no level crossing occurs
throughout the evolution; (ii) the derivatives of all the spectral
projectors $\{ |\psi_m(t)\rangle \langle \psi_m(t) | \}$ are
sufficiently smooth. The (normalized) time-evolved state reads
$$|\psi(t)\rangle = c_0 (t) |\psi_{0} (t)\rangle + \sum_{m \ne 0}
c_m(t) |\psi_m (t)\rangle,$$ \noindent for coefficients to be
determined.  Since for a truly adiabatic evolution no excitation is
induced in spite of the fact that the eigenstates of $H(t)$ evolve in
time, appropriately subtracting (following Berry, `renormalizing') the
adiabatic contribution is essential for quantifying the leading
non-adiabatic correction.  This is achieved in two steps
\cite{Messiah}: (i) Effect a canonical transformation to a `comoving
frame,' where in the zeroth-order adiabatic limit $\tau\rightarrow
\infty$ the comoving state vector $|\tilde{\psi}(t)\rangle =
\tilde{U}(t;t_{\text{in}}) |\psi(t_{\text{in}})\rangle$ is frozen up
to a phase factor, that is, $|\tilde{\psi}(t)\rangle = e^{-i \Gamma
_0(t) } |{\psi}_0 (t_{\text{in}})\rangle$, where $\Gamma_0(t)$
includes in general both the Berry phase and the dynamical phase;
%$\Gamma_0(t)= -i \int^t_{t_{\text{in}}} ds E_0(s)$ (in units $\hbar=1$);
(ii) Evaluate the first-order correction to the comoving-frame
propagator via Dyson series expansion.  Transforming back to the
physical frame, $c_m (t)= \langle \psi_m (t_{\text{in}}) |\tilde{U}
(t; t_{\text{in}}) | \psi_0 (t_{\text{in}}) \rangle$, to first-order
in $\dot{\lambda}$ we finally obtain (in units $\hbar=1$), $c_0^{(1)}
(t) = e^{-i\Gamma_0(t)} + O(\dot{\lambda}^2)$, and
\begin{eqnarray}
\hspace*{-4.5mm}c_m^{(1)}
(t)&\hspace*{-2.0mm}=\hspace*{-2.0mm}&e^{-i\Gamma\hspace*{-0.5mm}_m\hspace*{-0.5mm}(t)}
\hspace*{-2.0mm}\int_{t_{\text{in}}}^t \hspace*{-3.0mm} dt'
\dot{\lambda}(t') \frac{\langle
\psi_m(t')|H_1|\psi_0(t')\rangle}{E_m(t')-E_0(t')}\hspace*{0.5mm}
e^{i\int_{t_{\text{in}}}^{t'} \hspace*{-2.5mm} ds
\Delta_m(s) }  \hspace*{-3mm},  \nonumber \\
\hspace*{-4.5mm} \Delta_m(t) &\hspace*{-2.0mm} = \hspace*{-2.0mm}&
E_m(t)-E_0(t)\,. \label{tpt}
\end{eqnarray}

Knowledge of the time-dependent state enables arbitrary physical
quantities of interest to be computed, in particular the total
time-dependent excitation probability ${\mathbb P}_{\text{ex}}(t)=
\sum_{m \neq 0} |c_m (t)|^2$.  Given Eq.~(\ref{tpt}), the latter
formally recovers the expression given in \cite{Anatoli}, which
captures the contribution to the density of excitations from
states directly connected to $|\psi_0(t)\rangle$ via $H_1$
\footnote[2]{In particular, since one-body perturbations $H_1$ are
considered in the present analysis, the first-order excitation
probability, ${\mathbb P}_{\text{ex}}^{(1)}(t)= \sum_{m \neq 0}
|c_m^{(1)} (t)|^2$, coincides with the single-mode quasiparticle
contribution, $\langle N_{k,n}\rangle =1$, to the total
time-dependent excitation density.}. Dynamical scaling emerges
once the above result is supplemented by scaling assumptions on
three fundamental dynamical variables: the time-dependent
excitation energy above the ground state; the time-dependent
matrix elements of the perturbation; and the density of excited
states, $\rho(E)$, at the energy scale $\hat{t}^{-1}$
characterizing adiabaticity-breaking, which allows to change
discrete sums over excited states to integrals. That is, close to
the QCP we assume that:
\begin{eqnarray}
E_m (t)-E_0(t) & \hspace*{-1.8mm}=\hspace*{-1.8mm}&
\delta\lambda(t)^{\nu z}
f_m ({\Delta_m (t_c)}/{\delta\lambda(t)^{\nu z}}), \nonumber \\
\langle \psi_m (t)|H_1| \psi_0 (t)\rangle &\hspace*{-1.8mm} =
\hspace*{-1.8mm}& \delta\lambda(t)^{\nu
z-1} g_m ({\Delta_m (t_c) }/{\delta\lambda(t)^{\nu z}}), \nonumber \\
\rho(E) & \hspace*{-1.8mm}\sim \hspace*{-1.8mm}& E^{d/z-1},
\label{sassump1}
\end{eqnarray}
where the scaling functions $f_m, g_m$ satisfy i) $f_m$ ($g_m$) is
constant when $x \rightarrow 0$; ii) $f_m$ $(g_m)$ $\propto x$
when $x \rightarrow \infty$ \footnote[3]{ Note that $\rho \sim
\xi^{-d}/E$, with $\xi^d \sim \xi^m \xi^{d-m} \sim
E^{-m/z}L^{d-m}$ for a ($d-m$)-dimensional critical surface.}.
Having the scaling assumptions at hand, integration over excited
states is performed by moving to dimensionless variables $\zeta
=(t-t_c)/\hat{t}=(t-t_c) \tau^{-\nu z/(\nu z+1)}$ and $\eta=
\Delta_m (t_c) \hat{t}= \Delta_m (t_c) \tau^{\nu z/(\nu z+1)}$.
Since at the QCP the integrand in Eq.~(\ref{tpt}) develops a
simple pole, while the phase $e^{i\int_{t_{\text{in}}}^{t'}
\hspace*{-1.6mm} ds \Delta_m(s)}$ becomes stationary,
contributions away from the QCP may be neglected, allowing the
desired scaling factor to be isolated, up to a regular function
depending only on $\zeta$. Thus, the scaling of the excitation
density and {\em diagonal observables} such as the residual energy
is directly determined as $n_{\text{ex}}(\zeta) = \tau^{-d\nu/(\nu
z+1)} \Xi (\zeta), \; \Delta {H} (\zeta) = \tau^{-(d+z)\nu/(\nu
z+1)} \Upsilon (\zeta),$ see also Fig.~\ref{tpu}. For a {\em
generic observable}, if the additional scaling condition
\begin{equation}
\langle \psi_0(t)|\,{\cal O}\,|\psi_m(t) \rangle = \delta\lambda
(t)^\beta q_m({\Delta_m (t_c)}/{\delta\lambda(t)^{\nu z}}),
\label{sassump2}
\end{equation}
\noindent holds for all the excitations $m$ involved in the
process for an appropriate scaling function $q_m$, then $\Delta
{\cal O} \sim \tau^{-(\nu d +\beta)/(\nu z+1)}$ -- consistent with
Eq.~(\ref{scaling}).

Two remarks are in order.  First, the above argument directly
explains the dynamical scaling reported in \cite{deng} for
generalized entanglement relative to the fermionic algebra
$\mathfrak{u}(N)$ \cite{ge}, whose ground-state equilibrium
behavior directly reflects the fluctuations of the total number
operator.  Second, the derivation naturally extends to a generic
non-linear {\em power-law} quench, that is,
$\delta\lambda(t)=\lambda(t)-\lambda_c=|(t-t_c)/\tau|^\alpha
\text{sign}(t-t_c)$, $\alpha>0$.  Provided that the typical time
scale for adiabaticity breaking is redefined as $\hat{t}_\alpha
\sim \tau^{\alpha \nu z/ (1+\alpha \nu z)}$, the same scaling
assumptions in Eqs.~(\ref{sassump1})-(\ref{sassump2}) lead to
dynamical scaling behavior of the form $n_{ex} \sim \tau^{-\alpha
d\nu/(\alpha \nu z +1)}$, and $\Delta {\cal O} \sim
\tau^{-\alpha(d\nu+\beta)/(\alpha \nu z+1)}$, throughout the whole
time evolution \footnote[4]{ The perturbative derivation as
presented strictly applies to quenches across an isolated QCP
which is {\em not} multi-critical. We defer application of the
perturbative derivation to a multi-critical point to a forthcoming
analysis.}.

\begin{figure}
\begin{centering}
\includegraphics[width=\linewidth]{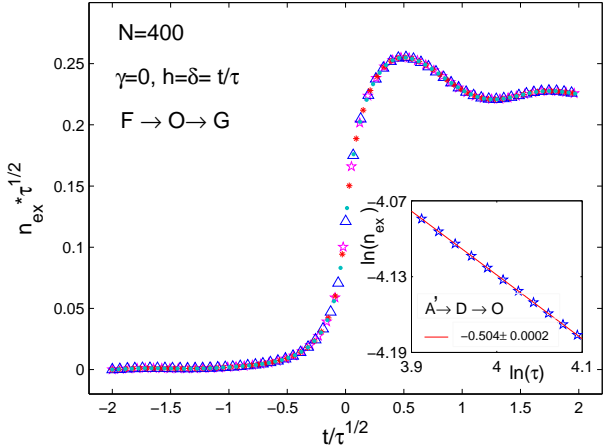}
\end{centering}
\vspace*{-5mm} \caption{\label{hd}(Color on-line) Main panel:
dynamical scaling of the excitation density for a simultaneous
linear quench of $h$ and $\delta$ along the gapless critical line
$F \rightarrow O \rightarrow G$. Inset: log-log plot of the final
excitation density vs $\tau$ along the path $A' \rightarrow D
\rightarrow O$.}
\end{figure}

%%%%%%%%%%%%%%%%%%%%%%%%%%%%%%%%%%%%%%%%%%%%%%%%%%%%%%%%%%%%%%%%%%%
\vspace*{0.5mm} {\bf Quenching along paths involving a finite
number of critical modes.--} A first situation which is beyond the
standard KZS discussed thus far arises in quenches that force the
system along a critical line, yet are dominated by a finite number
of participating excitations.  {\em Formally}, this makes it
possible to obtain the non-equilibrium exponent for
$n_{\text{ex}}$ through application of the KZS, provided care is
taken in defining the static exponents through a limiting
path-dependent process where, along the quench of interest, a {\em
simultaneous} expansion with respect to both the control parameter
{\em and} the relevant critical mode(s) is taken. Consider a
quenching scheme where both $h$ and $\delta$ are changed according
to $t/\tau$ while $\gamma=0$ (path $F \rightarrow O \rightarrow G$
in Fig. 1). While Eq.~(\ref{gap}) shows that the mode $k=\pi/2$ is
critical throughout the process ($\Delta_{\pi/2}(t) =0$, for all
$t \in [t_{\text{in}}, t_{\text{fin}}]$ as $N\rightarrow \infty$),
numerical data indicate that excitation sets in only when the
point $O$ is passed, see Fig.~\ref{hd}. As remarked, the static
critical exponents at $O$ are $z=1,\nu=1$, which differ from the
critical exponents ($z=2, \nu=1/2$) of all other critical points
along this line.  Indeed, the non-equilibrium exponent is solely
determined by the static exponents of this QCP along the chosen
path, $n_{\text{ex}} \sim \tau^{-\nu/(\nu z+1)}=\tau^{-1/2}$.  We
term a QCP which belongs to a different universality class than
all other critical points along a critical line and sets the
non-ergodic scaling a {\em dominant critical point for that line}
\footnote[5]{ The point $O$ is multi-critical. However, quenches
across a multi-critical point need {\em not} satisfy KZS, see
\cite{Sen2}.}. Physically, although $\Delta_{\pi/2}$ closes along
the critical line in the thermodynamic limit, a level crossing
which brings all bands together only occurs at $O$ -- still
allowing the time-evolved state to adiabatically follow the
snapshot ground state until then. The following independent
confirmations may be invoked in support of the above argument.
First, consider the anisotropic quench process analyzed in
\cite{Div}, whereby $\gamma(t)$ is changed linearly along the
critical line $h^2=\delta^2+1$.  By Taylor-expanding $\Delta_k$ in
Eq. (\ref{gap}) around $k=0, \gamma=0$ reveals that $\nu=1, z=2$
at the dominant QCP $(\gamma=0, h, \delta)$, whereas $\nu=1,z=1$
for $\gamma \ne 0$ along the line.  Accordingly, $n_{\text{ex}}
\sim \tau^{-1/3}$. While this coincides with the result obtained
in \cite{Div}, the underlying physical explanation is different.
Plots of the rescaled excitation density
($n_{\text{ex}}\tau^{1/3}$) vs the rescaled time ($t/\tau^{2/3}$)
would collapse onto one another for different $\tau$ within the
appropriate range, in complete analogy with Fig.~\ref{hd}. Second,
loss of adiabaticity at a single point can also explain the
scaling behavior observed for a AFM-to-FM quench (or a
critical-to-FM quench) in the XXZ model \cite{Pellegrini}, whereby
the control path involves the gapless critical region $-1 \leq
\Delta \leq 1$ and the dominant critical point $\Delta=1$ belongs
to a different universality class. Lastly, the concept of a
dominant QCP remains useful for a power-law quench, which leads to
the scaling behavior $n_{ex} \sim \tau^{-\alpha d\nu/(\alpha \nu z
+1)}$, with $\nu$ and $z$ being the critical exponents of the
dominant QCP along the critical line.

%%%%%%%%%%%%%%%%%%%%%%%%%%%%%%%%%%%%%%%%%%%%%%%%%%%%%%%%%%%%%%%%%%%%
\vspace*{0.5mm} {\bf Quenching along paths involving an infinite
number of critical modes.--} More complex scenarios emerge when
uncountably many modes of excitations can compete during the
quench. Focusing on the isotropic limit $\gamma=0$, we contrast
two representative situations where the Lifshitz QPT is involved:
(I) Magnetic quenches along the path $D \rightarrow O \rightarrow
E$ (PM $\rightarrow$ SF $\rightarrow$ PM); (II) Alternating
quenches along the path $A \rightarrow B \rightarrow C$ (DM
$\rightarrow$ SF $\rightarrow$ DM). Since $[M_z, H]=0$, in both
cases the allowed excitation must comply with a non-trivial
dynamical constraint. Along the path $D \rightarrow O \rightarrow
E$, this forces the final state to be the same as the initial
ground state up to a global phase factor, leading to
$n_{\text{ex}}(t_{\text{fin}}) \sim \tau^0$.  Although for a
magnetic quench this may be viewed as a consequence of the fact
that the dynamics simply acts as a relabelling of the snapshot
eigenstates, the same scaling holds for any quench which begins or
ends in the gapless phase -- for instance, a $\delta$-quench along
the path $A \rightarrow B$. Because these quenches take the system
through a critical line in momentum space, $d-m=1$ (as opposed to
$d-m=0$ for an isolated QCP), the observed scaling is consistent
with the recent prediction $n_{\text{ex}}(t_{\text{fin}})\sim
\tau^{-m \nu/(\nu z +1)}$ \cite{Sengupta}.

One may naively expect the same scaling to hold for path (II), which
also connects two gapped phases, albeit different than in (I). Unlike
in the standard KZS, however, details about the initial and final
phases as well as the time-dependent excitation pattern become
important.  Specifically, along path (II) we find
$n_{\text{ex}}(t_{\text{fin}}) \sim \tau^{-1/2}$. An explanation may
be obtained by exploiting the fact that due to $U(1)$-symmetry, the
fermion number is conserved.  This allows the reduced $4 \times 4$
matrix $H_k$ to be decoupled into two $2 \times 2$ matrices by
interchanging the order of the basis vectors $a_{-k}$ and
$b^\dag_k$. Thus, $\hat{H}_{\pm k}={W}^\dagger_{\pm k} {H'}_{\pm k}
{W}_{\pm k}$, where ${W}^\dagger_k=(a_k^\dagger, b_k^\dagger)$,
${W}^\dag_{-k}=(a_{-k}, b_{-k})$, and
\begin{eqnarray}
{H'}_{\pm k} = \pm 2h {\mathbb I}_2 + \left ( \begin{array}{cc}
\pm 2\delta & \mp2\cos k \\ \mp 2\cos k & \mp 2\delta
\end{array} \right). \label{twobytwo}
\end{eqnarray}
\noindent For such a two-level system, the asymptotic excitation
probability may be computed from the Landau-Zener transition
formula \cite{LZ}, yielding $p_k=e^{-2\pi \cos^2k \tau}$.  Upon
integrating over all modes, we find
\begin{eqnarray}
n_{\text{ex}}(t_{\text{fin}})= \frac{1}{\pi}\int_{-\pi/2}^{\pi/2}
dk \, p_k \sim \tau^{-1/2}. \label{LZ}
\end{eqnarray}
Note that because $p_k$ is independent on $h$, the result in Eq.
(\ref{LZ}) may be interpreted as implying that traversing the
gapless phase produces the same excitation density as crossing the
single QCP $O$ by translating path (II) at $h=0$, which determines
the non-equilibrium exponent.

\begin{figure}[t]
\begin{centering}
\includegraphics[width=\linewidth]{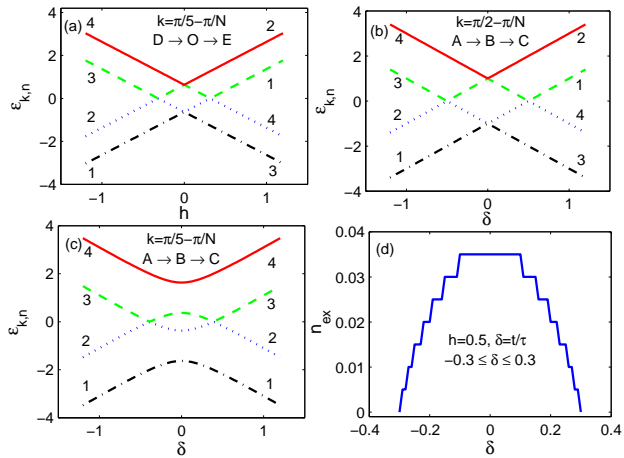}
\end{centering}
\vspace*{-5mm} \caption{\label{band} (Color on-line) Panels (a ---
c): band structure for different momentum modes $k$ vs quench
parameter ({\em i.e.}, $h$ or $\delta$).  Band ordering is
determined by band index $n=1,2,3,4$ at initial time
$t=t_{\text{in}}$, whereas dash-dotted (black) $\leq$ dotted
(blue) $\leq$ dashed (green) $\leq$ solid (red) at a generic time
$t$ along the control path. Panel (d): final excitation density,
$n_{\text{ex}}(t_{\text {fin}})$, vs $\delta$ for a quench of the
alternation strength $\delta$ at fixed $h=0.5$ within the critical
region.  For all cases, $N=400$.}
\end{figure}

Physical insight into what may be responsible for the different
behavior observed in the two quenches is gained by looking at the
excitation spectrum along the two paths.  Notice that, once the
energy eigenvalues are specified at the initial time,
($\epsilon_{k,1}(t_{\text{in}})\leq \epsilon_{k,2}(t_{\text{in}})
\leq\epsilon_{k,3}(t_{\text{in}})\leq\epsilon_{k,4}(t_{\text{in}})$
in our case), the same relative ordering need not hold at the
final time if a level crossing is encountered during the quench --
see Fig.~\ref{band}(a),(b).
%which means this relation $\epsilon_{k,1}(t_{\text{f}})\leq
%\epsilon_{k,2}(t_{\text{f}})\leq\epsilon_{k,3}(t_{\text{f}})\leq
%\epsilon_{k,4}(t_{\text{f}})$
In the critical region, a pair of modes $(k,n)$ and $(k,n')$
undergoes a level crossing if $h^2-\delta^2=\cos^2{k}$.  If the
number of such level crossings for fixed $n,n'$ is {\em even}, the
net contribution to the final excitation from momentum $k$ is
zero, since the final occupied bands are the same as the initial
ones -- see Fig.~\ref{band}(c).  No cancellation is in place if
either an odd number of level crossings from the same pair or if
different pairs $(n,n')$ are involved.  The latter situation is
realized for all $k$ along path (I) ($h$-quench,
Fig.~\ref{band}(a)) and also for the path $A \rightarrow B $
($\delta$-quench). For a $\delta$-quench along path (II), the net
excitation from the gapless phase turns out to be completely
canceled (as seen in Fig.~\ref{band}(d), where the quench starts
and ends symmetrically within the gapless phase). This only leaves
the two boundary critical lines $h=\pm\delta$ as contributing to
the excitation, thus a finite set of critical modes (a single one
in fact, $k=\pi/2$, see Fig.~\ref{band}(b)).
%for not able to applying $\tau^{-m\nu/(\nu z+1)}$ to
%$n_{\text{ex}}$ in path ({\tt B}).
Interestingly, a similar {\em cancellation mechanism} was verified
for repeated quenches across an isolated QCP \cite{Victor}.  While
a thorough analysis is beyond our current scope, we suggest that
even participation from the same (pair of) snapshot excitations
may be at the root of this cancellation in both scenarios. Here,
we further test this conjecture by examining the path $A'
\rightarrow D \rightarrow O$, for which an effective two-level LZ
mapping is no longer possible. Unlike $A \rightarrow B \rightarrow
C$, two intermediate phases are now crossed, and the initial and
final phase differ from one another, yet analysis of
$\epsilon_{k,n}(t)$ reveals that the two paths are equivalent in
terms of participation of critical modes. Numerical results
confirm that $n_{\text{ex}}(t_{\text{fin}}) \sim \tau^{-1/2}$,
Fig.~\ref{hd}.

\vspace*{0.5mm} {\bf Conclusions.--} Non-ergodic dynamical scaling
is fully captured by first-order adiabatic renormalization for
sufficiently slow quenches involving a simple isolated QCP. Beyond
this regime, we find that non-equilibrium exponents remain
expressible by combinations of equilibrium (path-dependent) ones
in all the scenarios under examination, however a detailed
characterization of both the static phase diagram and the
accessible low-lying excitations is necessary for quantitative
predictions.  Ultimately, scaling behavior appears to be the same
for control paths which share an equivalent excitation structure.
While yet different non-ergodic scaling may arise in more complex
systems ({\em e.g.}, infinite-order
Berezinskii-Kosterlitz-Thouless QPTs \cite{Pellegrini} as well as
models with infinite coordination \cite{LMG}), a deeper analysis
of how competing many-body excitations contribute and interfere
during a quench may shed further light on non-equilibrium quantum
critical physics. \vspace*{-2.5mm} \acknowledgments
\vspace*{-3.0mm} S.D. acknowledges partial support from Constance
and Walter Burke through their Special Projects Fund in QIS.
\vspace*{-3.5mm}

%%%%%%%%%%%%%%%%%%%%%%%%%%%%%%%%%%%%%%%%%%%%%%%%%%%%%%%%%%%%%%%%%%%%%%

\end{document}